# mRadNet: A Compact Radar Object Detector with MetaFormer

Huaiyu Chen*, Fahed Hassanat*†, Robert Laganière*†, Martin Bouchard*
*School of Electrical Engineering and Computer Science, University of Ottawa, Canada
†Sensor Cortek Inc., Canada

***Abstract*—Frequency-modulated continuous wave radars have gained increasing popularity in the automotive industry. Their robustness against adverse weather conditions makes it a suitable choice for radar object detection in advanced driver assistance systems. These real-time embedded systems have requirements for the compactness and efficiency of the model, which have been largely overlooked in previous work. In this work, we propose mRadNet, a novel radar object detection model with compactness in mind. mRadNet employs a U-net style architecture with MetaFormer blocks, in which separable convolution and attention token mixers are used to capture both local and global features effectively. More efficient token embedding and merging strategies are introduced to further facilitate the lightweight design. The performance of mRadNet is validated on the CRUW dataset, improving state-of-the-art performance with the fewest parameters and the lowest FLOPs.**

**Source code: https://github.com/huaiyu-chen/mRadNet.**

## I. Introduction

With recent advances in advanced driver assistance systems (ADAS), Frequency Modulated Continuous Wave (FMCW) radar has gained increasing popularity in the automotive industry [1]. Its ability to provide accurate distance and velocity measurements makes it a suitable choice for radar object detection (ROD) tasks. Since it operates in the millimeter-wave frequency range, FMCW radar is also less susceptible to adverse weather conditions such as rain, fog, and snow compared to traditional optical sensors [2]. This robustness is particularly important for ensuring reliable performance in various environmental conditions.

Data collected from FMCW radar sensors is typically preprocessed with 3D Fast Fourier Transforms (FFTs) into radio frequency (RF) images known as RAD data cubes, containing range, azimuth, and Doppler information. While rich in information, RAD cubes are difficult to interpret semantically. Efforts have been made to convert RAD data cubes into point clouds, which are often too sparse and noisy to be used for object detection tasks [3]. Other approaches process radar data in other modalities such as raw ADC data [4], but most recent works have focused on processing transformed RF images.

Convolutional neural networks (CNN) have seen widespread success in computer vision tasks. Many works have adopted CNN-based architectures for ROD tasks, where RF images are transformed from raw radar data, and fed into CNNs for object detection [5], [6], [7]. CNNs have been proven effective in extracting local features from radar data, but they often struggle to capture global dependencies due to their limited receptive fields. Small CNN models also fail to fully capture enough high level features from RF images, requiring CNN ROD models to be large and computationally expensive to achieve acceptable performance [3].

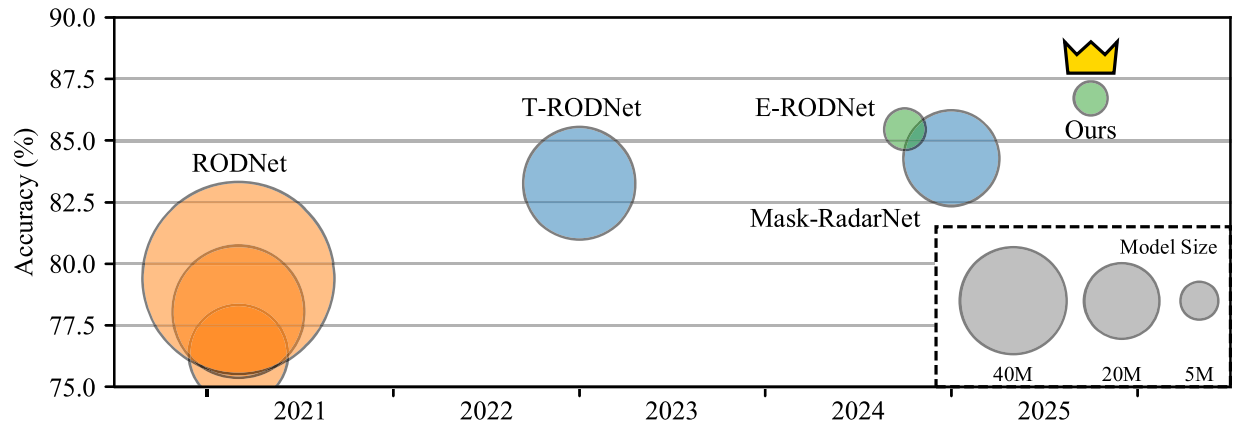

Fig. 1. Radar object detection accuracy on the CRUW dataset for CNNs, Transformers, and MetaFormers. We show that MetaFormers can outperform CNNs and Transformers in ROD tasks, while maintaining a compact design.

With the recent success of Transformer-based models in computer vision tasks, many works have started to explore the use of Transformers for ROD tasks [4], [8]. 3D Transformers have also seen usage in ROD tasks to process time-series data [9], [10], [11], allowing the model to capture temporal features and compensate for the noise in the data. Various mechanisms have been introduced to enhance the performance of Transformers in computer vision tasks, such as the use of sliding windows [12], but the high computational cost of the self-attention mechanism of Transformers remains unresolved.

Attributing the success of Transformers to their overall architecture rather than their attention modules, MetaFormers have set new records in various computer vision tasks [13], [14], [15], [16]. By replacing the self-attention module with simple token mixers, MetaFormers effortlessly achieve better performance than Transformers with significantly fewer parameters. Inspired by the success of MetaFormers, work has been done to apply MetaFormers to ROD tasks [17], but the use of all-convolutional layers has limited the ability to capture long-range dependencies, and the performance of MetaFormers on ROD tasks has not been fully explored.

To further exploit MetaFormer's potential in ROD tasks, we propose a novel architecture coined mRadNet. mRadNet benefits from the combination of convolutional and attention token mixers, allowing the model to capture both local and global features while maintaining a lightweight design. A U-Net [18] style overall architecture is adopted to produce a hierarchical representation, enabling the model to effectively capture both fine-grained details and high-level semantic information. Furthermore, more efficient token embedding and token merging strategies are introduced to cut down the computational cost of the model, paving the path for a deeper architecture with fewer parameters. All of these design choices form a lightweight and efficient architecture that improves state-of-the-art (SOTA) performance on the CRUW dataset [3]. Our contributions can be summarized as follows:

- We propose mRadNet, a novel MetaFormer-based, U-net style architecture for ROD tasks. MetaFormer blocks with convolutional layers are used to take the place of traditional CNN backbones and heads, forming a concise and cohesive design that captures both local and global features effectively.
- We introduce more efficient token embedding and token merging strategies to facilitate the effectiveness of the token mixers and the model's compactness.

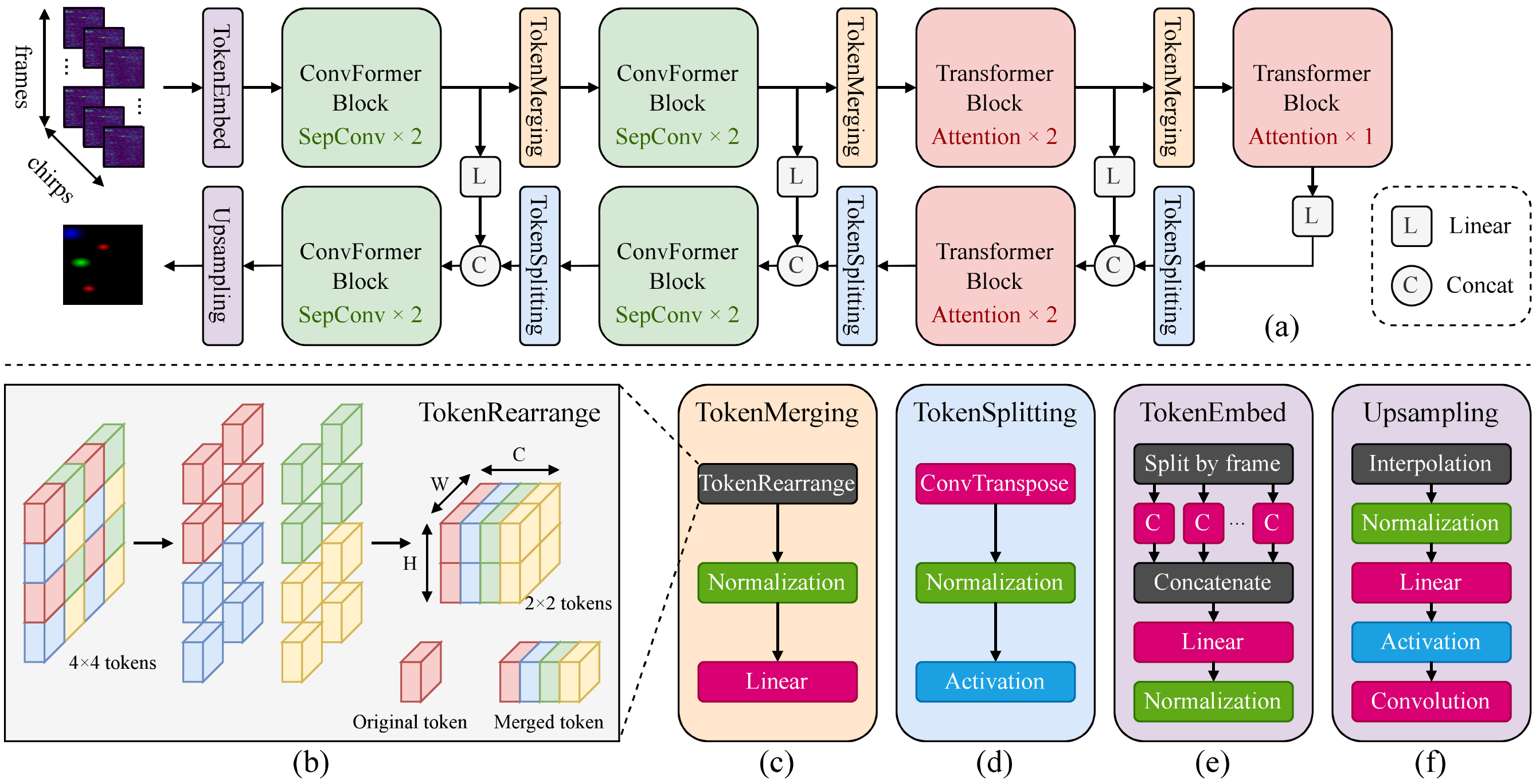

Fig. 2. Overall architecture of mRadNet.

The rest of the paper is organized as follows: section II describes the methodology, section III presents the experiments and results, and section IV concludes the paper.

## II. Methodology

### A. Overall Architecture

The overall architecture of mRadNet is shown in Figure 2 (a). mRadNet adopts a U-Net style architecture, which consists of an encoder and a decoder with skip connections between them. The network takes a batch of 5D tensors as input, consisting of multiple frames of complex RF images, each with several chirps, the real and imaginary components are treated as separate channels. Before entering the encoder, the input RF image is first passed through a stem, which extracts short-term temporal features and converts the complex RF image to a real-valued 3D image. In the encoder, the input is first tokenized with a TokenEmbed module, as described in subsection II-C. The tokens are then passed through a series of MetaFormer and Transformer blocks for hierarchical extraction of multilevel features. After each stage, a TokenMerging module is applied to form a hierarchical representation of the input, similar to a Swin Transformer design [12]. Each group of $2\times2$ neighboring tokens is merged into a single token, reducing the spatial resolution while increasing the information density. Note that the tokens are only merged in the spatial dimensions, tokens from different frames remain separated. The output of each MetaFormer block is stored for later use in the decoder.

In the decoder, another sequence of Transformer and MetaFormer blocks is applied to the tokens extracted from the encoder. Before each stage, a transposed convolution layer acts as the core of the token splitting module. With a kernel size of $1\times2\times2$ and a stride of $1\times2\times2$, each token is mapped to a $2\times2$ neighborhood of 4 new tokens. This module acts as the inverse of the TokenMerging module in the encoder, but the working mechanism is different since we cannot make assumptions about the semantic meaning of each element in the token. Thus, a ConvolutionTranspose layer with learnable parameters is used to learn the best way to split the tokens. The tokens are then concatenated with the corresponding tokens from the encoder through skip connections, before being passed through each stage. The final upsampling of the tokens is done through linear interpolation to ensure a smooth confidence map output, and avoid false positives caused by noisy local peaks. A final convolutional layer maps the upsampled tokens to the final output, i.e., confidence maps for each object class.

### B. Token Mixing

It has become common practice for Detection Transformers to prepend a convolutional backbone to Transformer encoders

[19], [20], [21]. This design choice stems from the fact that while Transformers excel at modeling global dependencies, they often struggle to capture local features effectively as compared to convolutional layers. MetaFormer, meanwhile, has pointed to an alternative direction of using other token mixers in place of the self-attention module, while retaining the overall framework.

Inspired by these works, mRadNet integrates the convolutional backbone into the encoder by prepending MetaFormer blocks with convolutional token mixers to the Transformer-based encoder. The same design is also applied to the end of the Transformer decoder, where convolutional token mixers are used to refine the local features in the upsampled tokens, and ensure a smooth output. Symmetry is maintained in the architecture, and skip connections found in U-net architectures also help to retain local features, by directly passing low-level features from the encoder to the decoder. This design forms a more cohesive architecture that effectively captures both local and global features. Each MetaFormer block can thus be described by Equation 1.

$$\begin{aligned} \boldsymbol{x}' &= \boldsymbol{x} + \mathrm{TokenMixer}(\mathrm{Norm}(\boldsymbol{x})) \\ \boldsymbol{y} &= \boldsymbol{x}' + \mathrm{Linear}(\mathrm{Norm}(\boldsymbol{x}')) \end{aligned} \tag{1}$$

where $\boldsymbol{x}$ is the input tokens, and $\boldsymbol{y}$ is the output tokens. The **TokenMixer** is responsible for information exchange between tokens, and the **Linear** layer is responsible for information exchange between channels. Standard convolution operations come with built-in channel mixing capabilities, which can be redundant when used in conjunction with the linear layer. To this end, we adopt depthwise separable convolution (SepConv) [22] as the convolutional token mixer in mRadNet. SepConv decomposes standard convolution into a depthwise convolution followed by a pointwise convolution, where depthwise convolution operates on each channel separately, and pointwise convolution is essentially a $1 \times 1 \times 1$ convolution that mixes information between channels.

### C. Token Embed Module

The CRUW dataset [3] offers radar data in the form of complex RA (range-azimuth) images for every chirp, each image with a shape of $H \times W \times 2$, where $H$ and $W$ are the number of range bins and azimuth bins and the last dimension represents the real and imaginary components of the complex image. Only 4 (0, 64, 128, and 192) out of 256 chirps in every frame are provided in the dataset, and each frame is sampled at 30 Hz. In order to compensate for the noise in the data and strengthen the temporal features, mRadNet processes multiple frames of radar data at once. The input to the model is a batch of 6D tensors with the shape of $B \times T \times N_c \times H \times W \times 2$, where $B$ is the batch size, $T$ is the number of frames, and $N_c$ is the number of chirps in each frame, which is 4 for the CRUW dataset.

The TokenEmbed module is designed to convert the input into a 3D array of tokens, which are then processed by the MetaFormer blocks. Since FMCW radar is a time-invariant system at the frame level, each frame can be treated equally. A single convolutional layer with a kernel size of $4 \times 1 \times 1$ and a stride of $4 \times 1 \times 1$ is applied on the (chirps, H, W) dimension, which effectively merges the chirps in every frame, and increases the number of channels. This operation is equivalent to applying a fully connected layer on the chirps dimension, which mimics the behavior of FFT in the preprocessing stage, but with learnable parameters. The chirps dimension is then removed, and the resulting tensor is reshaped to a 5D tensor (an RF "video") with the shape of $B \times T \times H \times W \times C$, where $C$ is the number of channels after the convolutional layer. The image is then split into non-overlapping patches, and each patch is represented by a single token, in line with ViT's design [23]. A 3D convolutional layer with a kernel size of $2 \times 2 \times 2$ and a stride of $2 \times 2 \times 2$ is applied to each patch on the (T, H, W) dimension. This layer groups every 8 neighboring pixels in the spatiotemporal dimensions into a single token, forming a $T/2 \times H/2 \times W/2$ array of tokens. We assume that these tokens are location-aware, and the following MetaFormer blocks with SepConv token mixers will help to retain the spatiotemporal information in the tokens.

### D. Token Merging Module

Token Merging modules are used in the encoder to downsample the tokens between stages. Instead of using pooling or strided convolution, we adopt a more structured approach by first rearranging neighboring tokens into new tokens with a Token Rearrange layer, as shown in Figure 2 (c). This operation is similar to the PixelShuffle operation commonly used in super-resolution tasks [24], but works in the opposite direction. Given an input tensor of size $B \times T \times H \times W \times C$, where $H$ and $W$ are the height and width of the feature map respectively, the Token Rearrange layer with a downscaling factor of 2 rearranges each non-overlapping $2 \times 2$ spatial neighborhood into a new token, resulting in an output tensor of size $B \times T \times \frac{1}{2}H \times \frac{1}{2}W \times 4C$. This operation effectively reduces the spatial dimensions by a factor of 4 while increasing the token dimension by the same factor, preserving all information from the input. Note that the temporal dimension remains unchanged. After the rearrangement, a normalization layer and a linear layer are applied to stabilize the training and reduce the token dimension back to $2C$, completing the downsampling process.

## III. Experiments

### A. Dataset

Our model is trained and evaluated on the CRUW dataset [3], a large-scale automotive radar object detection dataset containing 400K frames of camera-radar data sampled at 30Hz, of which a subset of 41K annotated frames (40 sequences) and 11K unannotated frames (10 sequences) is made public through the ROD2021 challenge. We divide the annotated subset into training and testing sets, with a ratio of 9:1.

### B. Evaluation Metrics

We evaluate the performance of our model with Object Location Similarity (OLS) [3], in line with other works evaluated on the dataset. OLS is defined in Equation 2:

$$\mathrm{OLS} = \exp\left(\frac{-d^2}{2(s\kappa_{cls})^2}\right) \tag{2}$$

TABLE I
COMPARISON OF MODEL PERFORMANCES ON CRUW DATASET

| Model | Overall | | Pedestrian | | Cyclist | | Car | | Params. (M) | FLOPs (G) | Latency (ms) | Training (h) |
|---|---|---|---|---|---|---|---|---|---|---|---|---|
| | AP (%) | AR (%) | AP (%) | AR (%) | AP (%) | AR (%) | AP (%) | AR (%) | | | | |
| RODNet-CDC[3] | 76.33 | 79.28 | 77.11 | 79.64 | 69.39 | 70.02 | 82.91 | 89.13 | 34.52 | 280.03 | **1** | **5** |
| RODNet-HG[3] | 79.43 | 83.59 | 78.90 | 83.81 | 76.69 | 78.85 | 83.36 | 88.55 | 129.19 | 2144.86 | 7 | 35 |
| RODNet-HGWI[3] | 78.06 | 81.07 | 79.47 | 81.85 | 70.35 | 71.40 | 84.39 | 90.05 | 61.29 | 5949.68 | 5 | 177 |
| T-RODNet[9] | 83.27 | 86.98 | 82.19 | 85.41 | 82.28 | 84.30 | 86.22 | 92.53 | 44.31 | 182.53 | 20 | 8 |
| Mask-RadarNet[10] | 84.29 | 87.36 | 82.74 | 85.80 | 85.06 | 86.67 | 85.96 | 90.66 | 32.12 | 176.91 | 20* | 8* |
| E-RODNet[17] | 85.46 | 89.19 | **84.30** | 88.37 | **88.56** | 90.95 | 83.87 | 90.14 | 6.10 | **33.25** | 36 | 6 |
| mRadNet (Ours) | **86.72** | **91.18** | 84.00 | **88.60** | 87.13 | **91.93** | **90.59** | **94.17** | **4.93** | **32.79** | 9 | **5** |

**Algorithm 1** Location-based NMS (L-NMS)

**Require:** $\mathcal{P}$ ▷ *Set of candidate peaks*
**Ensure:** $\mathcal{P}^*$ ▷ *Set of final predicted objects*
$\mathcal{P}^* \leftarrow \emptyset$
**while** $\mathcal{P} \neq \emptyset$ **do**
  $p \leftarrow \max(\mathcal{P})$ ▷ *Find the most confident peak*
  $\mathcal{P}^* \leftarrow \mathcal{P}^* \cup \{p\}, \mathcal{P} \leftarrow \mathcal{P} - \{p\}$ ▷ *Move $p$ to final set*
  **for** $p' \in \mathcal{P}$ **do**
    **if** $\mathrm{OLS}(p, p') > threshold$ **then**
      $\mathcal{P} \leftarrow \mathcal{P} - \{p'\}$ ▷ *Remove $p'$ from candidates*

where $d$ is the distance between the two objects, $s$ is the distance between the reference object and the radar, and $\kappa_{cls}$ is the manually tuned error tolerance for class $cls$. For each frame, location-based NMS (L-NMS) is applied to the predicted confidence maps to find the candidate detections, as described in Algorithm 1. The Hungarian algorithm is then applied to match the predictions with the ground truth, and OLS is calculated for each matched pair.

### C. Implementation Details

In the training process, a window size of 16 frames is used, and only the output confidence maps of the last 4 frames are used for evaluation. The window shifts by 4 frames at a time to cover the entire sequence. This strategy minimizes the delay between the frame arrival and the prediction — an aspect overlooked in previous works.

The model is trained with the AdamP optimizer [25], with the learning rate initialized to $1 \times 10^{-4}$ and decayed towards 0 with a cosine decay schedule. The model is trained for 20 epochs with a batch size of 4, taking around 5 hours on a single Nvidia RTX 4090 GPU. The loss is calculated with the Smooth L1 loss, which is defined in Equation 3:

$$\ell(x, y) = \begin{cases} 0.5(x-y)^2, & \text{if } |x-y| < 1 \\ |x-y| - 0.5, & \text{otherwise} \end{cases} \tag{3}$$

where $x$ and $y$ are the predicted and ground truth confidence maps, respectively.

### D. Comparison with SOTAs

To evaluate mRadNet's performance against existing methods, we compare its results with previous SOTA models that are evaluated on the same CRUW dataset. A full comparison with detailed class-wise results, as well as the model efficiencies in terms of the number of parameters, FLOPs, inference latency, and training time, is summarized in Table I.

mRadNet achieves the best performance in terms of both AP and AR, with 86.72% and 91.18% respectively, outperforming E-RODNet [17] by a margin of 1.26% and 1.99%. Notably, mRadNet achieves significant improvements in car detection, with more than 4% increase in AP compared to all other methods. This validates the effectiveness of the proposed mRadNet architecture in the detection of fast-moving objects through enhanced temporal modeling capability. For slow-moving objects, however, E-RODNet still holds a slight advantage in AP of 0.3% and 1.43% for pedestrian and cyclist, although mRadNet achieves better AR for both classes. This indicates that mRadNet's localization accuracy lags slightly behind E-RODNet for small and slow-moving objects, which could be attributed to E-RODNet's more extensive use of convolution operations that are more effective in capturing local features.

In terms of model efficiency, mRadNet achieves the best performance with the smallest model size and lowest computational cost among all compared methods. We observe a trend that MetaFormer-based models tend to be much smaller than both CNN and Transformer-based models, while achieving better performance as well. The MetaFormer architecture empowers these models to secure a solid performance lower bound, allowing more aggressive design optimizations to reduce model size and computational cost without sacrificing much accuracy. In terms of training and inference time, lightweight CNN models such as RODNet-CDC still hold an advantage due to their simple architectures, but their performance lags far behind more advanced models. If we rule out CNN models from the table, mRadNet also demonstrates the shortest training time and the lowest latency among Transformer and MetaFormer-based models, making it a well-balanced choice in both model performance and efficiency.

### E. Ablation Study

To better understand the contributions of the different components in mRadNet, we conduct an ablation study by replacing modules and evaluating the performance drop. The TokenEmbed module is replaced with average pooling layers and linear layers, the TokenArrange operation in the TokenMerging module is replaced with convolution with no normalization, and SepConv token mixers are replaced with standard Attention modules, respectively. The results show that by removing one of these components, the performance of the model drops significantly, highlighting the importance of each component in mRadNet. The results are summarized in Table II. We observe considerable performance drops in all three ablated models, with more than 3% decrease in overall AP and more than 1.5% decrease in overall AR. The model without SepConv token mixers suffers the most loss in localization accuracy, reflected by significant drops in pedestrian and overall AP, highlighting the importance of convolutional layers in capturing local features. The TokenEmbed module also plays a crucial role in temporal modeling, as its removal leads to a notable decrease in pedestrian and cyclist AP. Lastly, the model with replaced downsampling layers results in the lowest AR in all classes. Checkerboard effects caused by naive convolutional downsampling appear to have a detrimental impact on the model's ability to detect objects at various locations, leading to missed detections. Overall, the ablation study confirms the effectiveness of each component in mRadNet.

TABLE II
ABLATION STUDY ON MRADNET.

| Model | Overall | | Pedestrian | | Cyclist | | Car | |
|---|---|---|---|---|---|---|---|---|
| | AP% | AR% | AP% | AR% | AP% | AR% | AP% | AR% |
| -TokenEmbed | 83.31 | 88.56 | 80.95 | 84.97 | 82.94 | 91.36 | 87.55 | 91.25 |
| -TokenMerging | 83.07 | 86.43 | 81.38 | 84.61 | 83.34 | 85.11 | 85.55 | 90.86 |
| -SepConv | 82.89 | 89.14 | 78.84 | 85.92 | 83.36 | 90.44 | 88.97 | 92.90 |
| mRadNet | **86.72** | **91.18** | **84.00** | **88.60** | **87.13** | **91.93** | **90.59** | **94.17** |

## IV. CONCLUSION

In this paper, we propose mRadNet, a compact MetaFormer-based architecture for radar object detection tasks. By integrating convolutional and attention token mixers, mRadNet effectively captures both local and global features. Its U-Net-style design produces hierarchical representations that preserve fine-grained details while encoding high-level semantic information. Efficient token embedding and merging strategies reduce the computational cost and enable deeper architectures with fewer parameters. We validate mRadNet on the CRUW dataset, where it improves state-of-the-art performance with lower parameter count and FLOPs.

## REFERENCES

[1] L. Fan, J. Wang, Y. Chang, Y. Li, Y. Wang, and D. Cao, "4d mmwave radar for autonomous driving perception: A comprehensive survey," *IEEE Transactions on Intelligent Vehicles*, vol. 9, no. 4, pp. 4606–4620, 2024.

[2] A. Venon, Y. Dupuis, P. Vasseur, and P. Merriaux, "Millimeter wave fmcw radars for perception, recognition and localization in automotive applications: A survey," *IEEE Transactions on Intelligent Vehicles*, vol. 7, no. 3, pp. 533–555, 2022.

[3] Y. Wang, Z. Jiang, Y. Li, J. Hwang, G. Xing, and H. Liu, "Rodnet: A real-time radar object detection network cross-supervised by camera-radar fused object 3d localization," *IEEE Journal of Selected Topics in Signal Processing*, vol. 15, no. 4, pp. 954–967, 2021.

[4] J. Giroux, M. Bouchard, and R. Laganiere, "T-fftradnet: Object detection with swin vision transformers from raw adc radar signals," in *Proceedings of the IEEE/CVF International Conference on Computer Vision*, 2023, pp. 4030–4039.

[5] A. Zhang, F. Nowruzi, and R. Laganiere, "Raddet: Range-azimuth-doppler based radar object detection for dynamic road users," in *2021 18th Conference on Robots and Vision (CRV)*. IEEE, 2021, pp. 95–102.

[6] J. Rebut, A. Ouaknine, W. Malik, and P. Pérez, "Raw high-definition radar for multi-task learning," in *Proceedings of the IEEE/CVF Conference on Computer Vision and Pattern Recognition*, 2022, pp. 17 021–17 030.

[7] M. Sakthi, A. Tewfik, M. Arvinte, and H. Vikalo, "End-to-end system for object detection from sub-sampled radar data," in *2022 30th European Signal Processing Conference (EUSIPCO)*. IEEE, 2022, pp. 1886–1890.

[8] L. Cheng and S. Cao, "Transrad: Retentive vision transformer for enhanced radar object detection," *IEEE Transactions on Radar Systems*, 2025.

[9] T. Jiang, L. Zhuang, Q. An, J. Wang, K. Xiao, and A. Wang, "T-rodnet: Transformer for vehicular millimeter-wave radar object detection," *IEEE Transactions on Instrumentation and Measurement*, vol. 72, pp. 1–12, 2022.

[10] Y. Wu, J. Liu, G. Jiang, W. Liu, D. Orlando, and L. Xiao, "Mask-radarnet: Enhancing radar object detection with spatio-temporal context," *IEEE Transactions on Intelligent Transportation Systems*, vol. 27, no. 1, pp. 1039–1051, 2025.

[11] L. Zhuang, Y. Yao, and N. Li, "Rc-rosnet: Fusing 3d radar range-angle heat maps and camera images for radar object segmentation," *IEEE Transactions on Circuits and Systems for Video Technology*, 2025.

[12] Z. Liu, Y. Lin, Y. Cao, H. Hu, Y. Wei, Z. Zhang, S. Lin, and B. Guo, "Swin transformer: Hierarchical vision transformer using shifted windows," in *Proceedings of the IEEE/CVF International Conference on Computer Vision*, 2021, pp. 10 012–10 022.

[13] W. Yu, C. Si, P. Zhou, M. Luo, Y. Zhou, J. Feng, S. Yan, and X. Wang, "Metaformer baselines for vision," *IEEE Transactions on Pattern Analysis and Machine Intelligence*, vol. 46, no. 2, pp. 896–912, 2023.

[14] B. Kang, S. Moon, Y. Cho, H. Yu, and S. Kang, "Metaseg: Metaformer-based global contexts-aware network for efficient semantic segmentation," in *Proceedings of the IEEE/CVF Winter Conference on Applications of Computer Vision*, 2024, pp. 434–443.

[15] X. Chen, F. Ma, Y. Wu, B. Han, L. Luo, and S. Biancardo, "Mfmdepth: Metaformer-based monocular metric depth estimation for distance measurement in ports," *Computers & Industrial Engineering*, vol. 207, pp. 1–13, 2025.

[16] J. Kim and G. Kwon, "Efficient classification of photovoltaic module defects in infrared images," *IEEE Signal Processing Letters*, vol. 32, pp. 2389–2393, 2025.

[17] W. Xu, P. Lu, and Y. Zhao, "E-rodnet: lightweight approach to object detection by vehicular millimeter-wave radar," *IEEE Sensors Journal*, vol. 24, pp. 33 091–33 100, 2024.

[18] O. Ronneberger, P. Fischer, and T. Brox, "U-net: Convolutional networks for biomedical image segmentation," in *International Conference on Medical Image Computing and Computer-assisted Intervention*. Springer, 2015, pp. 234–241.

[19] N. Carion, F. Massa, G. Synnaeve, N. Usunier, A. Kirillov, and S. Zagoruyko, "End-to-end object detection with transformers," in *European conference on computer vision*. Springer, 2020, pp. 213–229.

[20] X. Zhu, W. Su, L. Lu, B. Li, X. Wang, and J. Dai, "Deformable detr: Deformable transformers for end-to-end object detection," *arXiv preprint arXiv:2010.04159*, 2020.

[21] Y. Zhao, W. Lv, S. Xu, J. Wei, G. Wang, Q. Dang, Y. Liu, and J. Chen, "Detrs beat yolos on real-time object detection," in *Proceedings of the IEEE/CVF conference on computer vision and pattern recognition*, 2024, pp. 16 965–16 974.

[22] F. Chollet, "Xception: Deep learning with depthwise separable convolutions," in *Proceedings of the IEEE conference on computer vision and pattern recognition*, 2017, pp. 1251–1258.

[23] A. Dosovitskiy, L. Beyer, A. Kolesnikov, D. Weissenborn, X. Zhai, T. Unterthiner, M. Dehghani, M. Minderer, G. Heigold, S. Gelly *et al.*, "An image is worth 16x16 words: Transformers for image recognition at scale," *arXiv preprint arXiv:2010.11929*, 2020.

[24] W. Shi, J. Caballero, F. Huszár, J. Totz, A. P. Aitken, R. Bishop, D. Rueckert, and Z. Wang, "Real-time single image and video super-resolution using an efficient sub-pixel convolutional neural network," in *Proceedings of the IEEE conference on computer vision and pattern recognition*, 2016, pp. 1874–1883.

[25] B. Heo, S. Chun, S. Oh, D. Han, S. Yun, G. Kim, Y. Uh, and J. Ha, "Adamp: Slowing down the slowdown for momentum optimizers on scale-invariant weights," *arXiv preprint arXiv:2006.08217*, 2020.